# Self-compressed inhomogeneous stabilized jellium model and surface relaxation of simple metal thin films

M. Payami[1] and T. Mahmoodi[2]

1. Physics Group, Nuclear Science and Technology Research Institute, AEOI, Tehran, Iran
2. Department of Physics, Faculty of Sciences, Mashhad Branch, Islamic Azad University, Iran

Email: mpayami@aeoi.org.ir

## Abstract

The interlayer spacings near the surface of a crystal are different from that of the bulk. As a result, the value of the ionic density in the normal direction and near to the surface shows some oscillations around the bulk value. To describe this behavior in a simple way, we have formulated the self-compressed inhomogeneous stabilized jellium model and have applied it to simple metal thin films. In this model, for a $v$-layered slab, each ionic layer is replaced by a jellium slice of constant density. The equilibrium densities of the slices are determined by minimizing the total energy per electron of the slab with respect to the slice densities. However, to avoid the complications that arise due to the increasing number of independent slice-density parameters for large-$v$ slabs, we consider a simplified version of the model which consists of only three jellium slices: one inner bulk slice with density $\bar{n}_1$ and two similar surface slices, each of density $\bar{n}_2$. In this simplified model, each slice may contain more than one ionic layer. Application of this model to the $v$-layered slabs ($3 \leq v \leq 10$) of Al, Na, and Cs shows that, in the equilibrium state, $\bar{n}_1$ differs from $\bar{n}_2$. The difference is significant in the Al case, and the slab is more stable than that predicted by the homogeneous model with only one density parameter for the whole jellium background. In addition, we have calculated the overall relaxations, the work functions, and the surface energies, and compared with the results of earlier works.

## Introduction

An atomic slab is a system composed of a finite number of atomic layers stacked over in the $z$ direction, and each layer is extended to infinity in the other two directions of $x$ and $y$. The finiteness of the size in the $z$ direction gives rise to the surface effects which cause the interlayer spacings near the surface become different from the bulk value. In our recent work [1], using the self-compressed stabilized jellium model (SC-SJM) [2-5], we had studied the equilibrium properties of Al, Na, and Cs slabs and obtained some good results comparable to those obtained from the first-principles calculations. In the SC-SJM slab calculations [1], the discrete ions of atoms were replaced by a uniform positive charge density $\bar{n}$, and it was shown that in the equilibrium state, $\bar{n}$, or equivalently the density parameter $r_s = (3/4\pi\bar{n})^{1/3}$, assumes values different from that of the bulk. However, the information carried by the single equilibrium value $\bar{n}$ allows us only to determine whether the slab is expanded or contracted relative to the corresponding slab as part of a bulk system, and is not enough to describe the details of relaxations in the interlayer spacings near the surface. In order to describe the relaxations in different regions of a slab, we have to distinguish between the behaviors of different regions (for example, the relaxations near the surface are significant while, they are negligible deep inside the bulk region.) The full description of this observation becomes possible when we release the constraint of using a single jellium density for the whole system, and instead, attribute different jellium densities for the different regions which lead to the formulation of the inhomogeneous stabilized jellium model (ISJM) that is the subject of this work. As in the SJM, the ISJM can be applied in two ways. In the first method, one can use the experimental interlayer spacings to determine the corresponding jellium densities at different regions and use them as input parameters. However, in the second method, which is the self-compressed version (SC-ISJM), the jellium density at each region is determined from the total energy minimization with respect to the jellium densities in all regions. That is, each region assumes a certain background density which makes the force acting on that region vanish. In this way, the mechanical stability governs each region of the whole system. The SC-ISJM can be applied, in principle, to a system with an arbitrary number of inequivalent regions. However, by increasing the number of inequivalent regions beyond two (where only two parameters $\bar{n}_1$ and $\bar{n}_2$ are needed), the number of configurations in parameter space increases dramatically, so that it becomes impossible to find the global minimum by simple methods, and one has to resort to the sophisticated methods of global minimization. In that case, the computational effort increases to the level of atomic simulation. However, to avoid such complications, and still keep the computational costs much less than those of first-principles atomic simulations, we distinguish three main regions in a slab: two equivalent surface regions and one inner region. The inner part, which is specified by a jellium density $\bar{n}_1$, contains those ionic layers that have more or less the bulk spacings (for sufficiently thick slabs); and the identical surface parts, which are specified by jellium density $\bar{n}_2$, contain a few surface layers having significant fluctuations in the interlayer spacings. In this work, we have applied the SC-ISJM to $\nu$-layered (100) slabs of Al, Na, and Cs with $3 \leq \nu \leq 10$. The results show that the slabs undergo significant relaxations in the surface regions as expected, while the relaxations in the bulk regions are negligibly small for sufficiently thick slabs. The calculations are based on the self-consistent solutions of the Kohn-Sham (KS) equations [6] in the density functional theory (DFT) [7] using the local density approximation (LDA) [6] for the exchange-correlation. Our results show that in the SC-ISJM, the slab systems are more stable than in the SC-SJM [1]. In section 2 we present the formulation of the general SC-ISJM. The calculation details are explained in section 3, and we have discussed the calculation results in section 4. To show how the properties of slabs evolve with their sizes, the self-consistent electron and jellium densities plots of some slabs are presented in the appendix at the end of the paper.

**Formulation of the SC-ISJM**

We start from the energy functional of a system of electrons of density $n(\vec{r})$ interacting with ions (distributed over sites $\{\vec{l}\}$) via a local pseudopotential, given by

$$E[n;\{\vec{l}\}] = T_s[n] + E_{xc}[n] + \frac{1}{2}\int d\vec{r}d\vec{r}' \frac{n(\vec{r})n(\vec{r}')}{|\vec{r}-\vec{r}'|} \qquad (1)$$
$$+ \int d\vec{r} \sum_{\vec{l}} w(|\vec{r}-\vec{l}|) n(\vec{r}) + \frac{1}{2}\sum_{\vec{l}\neq\vec{l}'} \frac{z^2}{|\vec{l}-\vec{l}'|}.$$

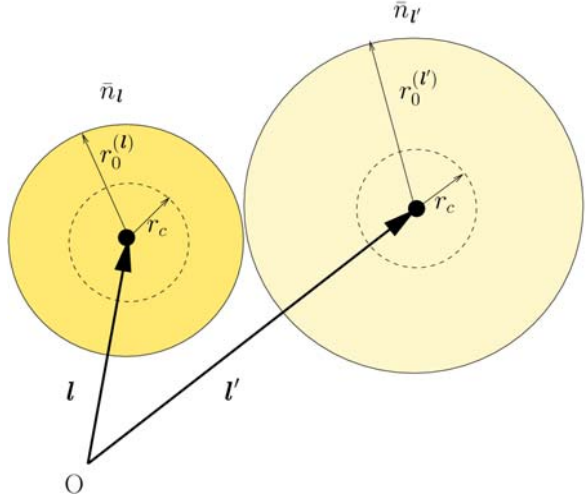

**Figure 1.** Jellium spheres centered at two nearest neighbor ionic sites $\vec{l}$ and $\vec{l}'$. The constant densities, $\bar{n}_{\vec{l}}$, and thereby, the radii, $r_0^{(\vec{l})}$, of the spheres at different sites may differ. The pseudopotential core radii, $r_c$, are the same at all sites.

The first three terms of equation (1) are the noninteracting kinetic, exchange-correlation, and Hartree energies, respectively; and the last two terms give the electron-ion and ion-ion interaction energies, respectively. All equations throughout this paper are expressed in Hartree atomic units. For the ionic pseudopotential $w$, the Ashcroft empty-core pseudopotential [8] is used which consists of a long-range attractive part and a short-range repulsive part $w_R$,

$$w(r) = -\frac{z}{r} + w_R(r); \quad w_R(r) = +\frac{z}{r}\theta(r_c - r), \qquad (2)$$

in which $z$ is the ionic charge, $\theta$ is the step function, and $r_c$ is the core radius of the pseudopotential which is determined from the bulk stability condition. We now consider a neutralizing positive charge distribution $n_+(\vec{r})$, which is composed of non-overlapping jellium spheres centered at ionic sites,

$$n_+(\vec{r}) = \sum_{\vec{l}} \bar{n}_{\vec{l}} \theta(r_0^{(\vec{l})} - |\vec{r} - \vec{l}|) \equiv \sum_{\vec{l}} n_+^{(\vec{l})}(\vec{r}). \tag{3}$$

Each sphere has its own radius, $r_0^{(\vec{l})} = z^{1/3} r_s^{(\vec{l})}$, and uniform density, $\bar{n}_{\vec{l}} = 3/4\pi [r_s^{(\vec{l})}]^3$ (figure 1). Adding and subtracting $n_+(\vec{r})$ to equation (1) yields [9]

$$\begin{aligned} E[n, n_+; \{\vec{l}\}] &= E_1[n, n_+] + \int d\vec{r}\, \delta v(\vec{r})[n(\vec{r}) - n_+(\vec{r})] \\ &+ \int d\vec{r} \sum_{\vec{l}} w_R(|\vec{r} - \vec{l}|) n_+(\vec{r}) \\ &+ \left\{ \frac{1}{2} \int d\vec{r} d\vec{r}'\, \frac{n_+(\vec{r}) n_+(\vec{r}')}{|\vec{r} - \vec{r}'|} + \frac{1}{2} \sum_{\vec{l} \neq \vec{l}'} \frac{z^2}{|\vec{l} - \vec{l}'|} - \int d\vec{r} \sum_{\vec{l}} \frac{z}{|\vec{r} - \vec{l}|} n_+(\vec{r}) \right\}, \end{aligned} \tag{4}$$

in which

$$E_1[n, n_+] = T_s[n] + E_{xc}[n] + \frac{1}{2} \int d\vec{r}\, \phi([n, n_+]; \vec{r})[n(\vec{r}) - n_+(\vec{r})], \tag{5}$$

with

$$\phi([n, n_+]; \vec{r}) = \int d\vec{r}'\, \frac{[n(\vec{r}') - n_+(\vec{r}')]}{|\vec{r} - \vec{r}'|}, \tag{6}$$

has the same functional structure as the energy in the ordinary jellium model (JM). In the second term of equation (4), the difference potential $\delta v$, defined by

$$\delta v(\vec{r}) = \sum_{\vec{l}} w(|\vec{r} - \vec{l}|) + \int d\vec{r}'\, \frac{n_+(\vec{r}')}{|\vec{r} - \vec{r}'|} \equiv \sum_{\vec{l}} \delta v^{(\vec{l})}(\vec{r}), \tag{7}$$

with

$$\delta v^{(\vec{l})}(\vec{r}) = w(|\vec{r} - \vec{l}|) + \int d\vec{r}'\, \frac{n_+^{(\vec{l})}(\vec{r}')}{|\vec{r} - \vec{r}'|}, \tag{8}$$

is the difference potential between the pseudopotential of ions and the electrostatic potential of the jellium spheres. The bracketed term in equation (4) is the electrostatic energy of a system composed of negative-charge jellium spheres and positive ions, in correspondence with the Madelung energy. To evaluate this "Madelung energy", in the first term of brackets we use the decomposition

$$n_+(\vec{r}) n_+(\vec{r}') = \sum_{\vec{l}} n_+^{(\vec{l})}(\vec{r}) n_+^{(\vec{l})}(\vec{r}') + \sum_{\vec{l} \neq \vec{l}'} n_+^{(\vec{l})}(\vec{r}) n_+^{(\vec{l}')}(\vec{r}'), \tag{9}$$

which leads to

$$\frac{1}{2} \int d\vec{r} d\vec{r}'\, \frac{n_+(\vec{r}) n_+(\vec{r}')}{|\vec{r} - \vec{r}'|} = \sum_{\vec{l}} z \tilde{\varepsilon}_{\vec{l}} + \frac{1}{2} \sum_{\vec{l} \neq \vec{l}'} \frac{z^2}{|\vec{l} - \vec{l}'|}, \tag{10}$$

in which

$$\tilde{\varepsilon}_{\vec{l}} = \frac{3z}{5r_0^{(\vec{l})}} \tag{11}$$

is the self-interaction energy per unit charge of the sphere at site $\vec{l}$, and the second term gives the sum of interactions between different spheres. On the other hand, the third term in the brackets reduces to

$$\int d\vec{r} \sum_{\vec{l}} \frac{z}{|\vec{r}-\vec{l}|} n_+(\vec{r}) = \sum_{\vec{l}} \frac{3z^2}{2r_0^{(\vec{l})}} + \sum_{\vec{l}\neq\vec{l}'} \frac{z^2}{|\vec{l}-\vec{l}'|}, \tag{12}$$

where the first term is the sum of interaction energies of spheres with their ions at the center, and the second term is the sum of interaction energies of point ions at $\vec{l}$ with spheres centered at $\vec{l}'$ ($\vec{l} \neq \vec{l}'$) which is twice the sum of sphere-sphere interaction energies. Summing up the results, the bracketed term in equation (4) reduces to

$$\{\cdots\} = \sum_{\vec{l}} z\varepsilon_M^{(\vec{l})}; \qquad \varepsilon_M^{(\vec{l})} = -\frac{9z}{10r_0^{(\vec{l})}}, \tag{13}$$

where $\varepsilon_M^{(\vec{l})}$ is the contribution of jellium sphere at site $\vec{l}$ to the "Madelung energy". Finally, assuming that the inequality $r_c < r_0^{(\vec{l})}$ holds for all sites (i.e., we do not have very high ionic density regions), the third term in equation (4) reduces to

$$\int d\vec{r} \sum_{\vec{l}} w_R(|\vec{r}-\vec{l}|) n_+(\vec{r}) = \sum_{\vec{l}} z\bar{w}_R^{(\vec{l})}, \tag{14}$$

where $\bar{w}_R^{(\vec{l})}$ is the average of the repulsive part of the pseudopotential over the sphere of volume $\Omega^{(\vec{l})} = (4\pi/3)[r_0^{(\vec{l})}]^3$ centered at site $\vec{l}$,

$$\bar{w}_R^{(\vec{l})} = \frac{1}{\Omega^{(\vec{l})}} \int d\vec{r}\, w_R(|\vec{r}-\vec{l}|) = 2\pi\bar{n}_{\vec{l}} r_c^2. \tag{15}$$

Inserting the results into equation (4), and using the identity

$$z = \int d\vec{r}\, n_+^{(\vec{l})}(\vec{r}), \tag{16}$$

we obtain

$$E[n, n_+; \{\vec{l}\}] = E_1[n, n_+] + \sum_{\vec{l}} \int d\vec{r}\, \delta v^{(\vec{l})}(\vec{r})[n(\vec{r}) - n_+(\vec{r})] \\ + \sum_{\vec{l}} (\bar{w}_R^{(\vec{l})} + \varepsilon_M^{(\vec{l})}) \int d\vec{r}\, n_+^{(\vec{l})}(\vec{r}). \tag{17}$$

It is also straightforward to show that the equality

$$<\delta v^{(\vec{l})}> \equiv \frac{1}{\Omega^{(\vec{l})}} \int d\vec{r}\, \delta v^{(\vec{l})}(\vec{r}) = \overline{w}_R^{(\vec{l})} + \tilde{\varepsilon}_{\vec{l}} + \varepsilon_M^{(\vec{l})}, \qquad (18)$$

holds for arbitrary $\vec{l}$. So far, equation (17) is equivalent to equation (1) and no approximation has been made. To go further and develop the ISJM, it can easily be shown that, taking the origin of coordinate system at site $\vec{l}$, the $\delta v^{(\vec{l})}(\vec{r})$ is given by

$$\delta v^{(\vec{l})}(r) = \begin{cases} w(r) + \dfrac{z}{r_0^{(\vec{l})}}\left(\dfrac{3}{2} - \dfrac{r^2}{2[r_0^{(\vec{l})}]^2}\right); & r < r_0^{(\vec{l})}, \\ 0 & ; \; r \geq r_0^{(\vec{l})}. \end{cases} \qquad (19)$$

At this point, we make our first approximation by taking the average value $\left\langle w(r) + \dfrac{z}{r_0^{(\vec{l})}}\left(\dfrac{3}{2} - \dfrac{r^2}{2[r_0^{(\vec{l})}]^2}\right)\right\rangle$ for $r < r_0^{(\vec{l})}$, and zero for $r \geq r_0^{(\vec{l})}$; which is equivalent to taking

$$\delta v^{(\vec{l})}(\vec{r}) \approx <\delta v^{(\vec{l})}> \theta(r_0^{(\vec{l})} - |\vec{r} - \vec{l}|). \qquad (20)$$

In the second stage of approximations, we replace the closely-packed jellium spheres with a nonuniform jellium background having sharp boundaries. Each part of the boundary separates two regions of different constant jellium densities, and the volume of a region is taken equal to the sum of volumes of the spheres in that region. Applying these two approximations to equation (17), we obtain the ISJM energy functional as

$$E_{ISJM}[n, \{n_+^{(\alpha)}\}] = E_{JM}[n, \{n_+^{(\alpha)}\}] + \sum_\alpha <\delta v>_{WS}(\overline{n}_\alpha) \int d\vec{r}\, \Theta^{(\alpha)}(\vec{r})[n(\vec{r}) - n_+^{(\alpha)}(\vec{r})]$$
$$+ \sum_\alpha [\overline{w}_R(\overline{n}_\alpha) + \varepsilon_M(\overline{n}_\alpha)] \int d\vec{r}\, n_+^{(\alpha)}(\vec{r}). \qquad (21)$$

In equation (21),

$$n_+^{(\alpha)}(\vec{r}) \equiv \overline{n}_\alpha \Theta^{(\alpha)}(\vec{r}), \qquad (22)$$

$$<\delta v>_{WS}(\overline{n}_\alpha) = \frac{1}{\Omega^{(\alpha)}} \int_{\Omega^{(\alpha)}} d\vec{r}\, \delta v(\vec{r}); \qquad \Omega^{(\alpha)} = \frac{z}{\overline{n}_\alpha}, \qquad (23)$$

where $\Theta^{(\alpha)}(\vec{r})$ takes the value of unity inside the region $\alpha$ and zero outside. $\overline{n}_\alpha$ is the constant jellium density of region $\alpha$. The equilibrium state of the system is determined by the minimization rule

$$E_0/N \equiv E(\{\overline{n}_\alpha^\dagger\})/N = \min_{\{\overline{n}_\alpha\}} E(\{\overline{n}_\alpha\})/N, \qquad (24)$$

where $N$ is the total number of valence electrons, $E(\{\bar{n}_\alpha\})$ is the ground-state energy of the system constrained to the external parameters $\{\bar{n}_\alpha\}$, and $\{\bar{n}_\alpha^\dagger\}$ is the set of external parameters that corresponds to the equilibrium state of the isolated $N$-electron system.

In applying the ISJM to a slab system, each atomic layer can be replaced by a jellium slice of corresponding constant density. However, as we discussed earlier, for slabs with more than a few atomic layers, the self-consistent determination of equilibrium densities of jellium slices becomes a difficult job and contradicts the simplicity of the model. To avoid such complications, for a $\nu$-layered slab we distinguish three main regions: one inner region containing $\nu_1$ atomic layers and two identical surface regions each containing $\nu_2$ surface atomic layers ($\nu_1 + 2\nu_2 = \nu$). The inner region, which is specified by a jellium density $\bar{n}_1$, contains those ionic layers that have more or less the bulk spacings (for sufficiently thick slabs); and the equivalent surface regions, which are specified by jellium density $\bar{n}_2$, contain a few surface layers having significant fluctuations in the interlayer spacings. The situation is schematically shown in figure 2. The equilibrium-state densities of the inner jellium, $\bar{n}_1^\dagger$ and the surface jellium, $\bar{n}_2^\dagger$, as well as the number of atomic layers in the surface region, $\nu_2^\dagger$, are determined self-consistently by the minimization rule

$$E_0(\nu)/N \equiv E(\nu, \nu_2^\dagger; \bar{n}_1^\dagger, \bar{n}_2^\dagger)/N = \min_{\nu_2}\{\min_{\bar{n}_1, \bar{n}_2} E(\nu, \nu_2; \bar{n}_1, \bar{n}_2)/N\}, \qquad (25)$$

where $E(\nu, \nu_2; \bar{n}_1, \bar{n}_2)$ is the ground-state energy of a $\nu$-layered slab with a given set of external parameters $\nu_2$, $\bar{n}_1$, and $\bar{n}_2$ which is calculated from the self-consistent solutions of the KS equations.

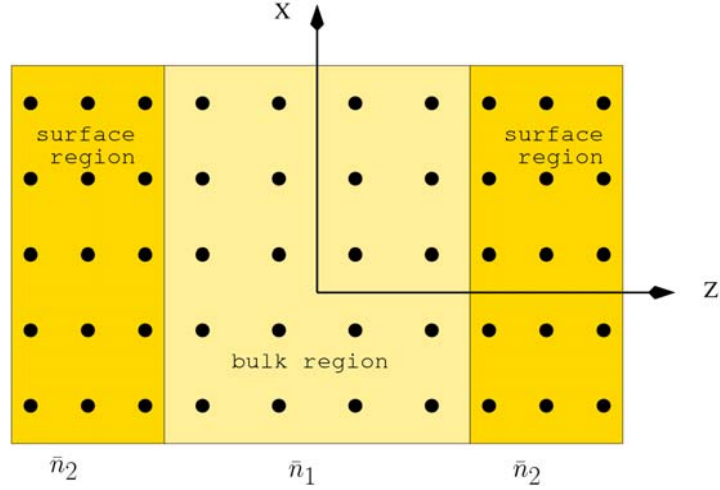

**Figure 2.** Schematic representation of the simplified ISJM for a slab with three main regions. The system is extended to infinity in the $x$ and $y$ directions. In this example, each surface region contains three atomic layers, while the inner bulk region contains four atomic layers. The interlayer spacings in the surface region can be different from that in the inner region.

**Calculation details**

In the ISJM, the slab has a full translational symmetry in the *x* and *y* directions, and because of finiteness in the *z* direction, the physical quantities depend on the *z* coordinate. Moreover, the KS equations reduce to one-dimensional equations,

$$\left(-\frac{1}{2}\frac{d^2}{dz^2} + v_{eff}(z)\right)\psi_m(z) = \varepsilon_m \psi_m(z), \quad m = 1, 2, \ldots \qquad (26)$$

where

$$v_{eff}(z) = \phi(z) + v_{xc}(z) + \sum_\alpha <\delta v>_{WS} (\bar{n}_\alpha) \Theta^{(\alpha)}(z). \qquad (27)$$

$\phi(z)$ is the electrostatic potential energy,

$$\phi(z) = 4\pi \int_{-\infty}^{z} dz' (z - z')[n_+(z') - n(z')], \qquad (28)$$

with

$$n_+(z) = \sum_\alpha \bar{n}_\alpha \Theta^{(\alpha)}(z), \qquad (29)$$

and satisfying the boundary condition $\phi(\pm\infty) = 0$. The electron density $n(z)$ is given by [10]:

$$n(z) = \frac{1}{\pi} \sum_{\varepsilon_m \leq E_F} (E_F - \varepsilon_m) \psi_m^2(z), \qquad (30)$$

where the Fermi energy is determined from the charge neutrality condition. In the simplified model, $\alpha$ takes the values 1, 2, and 3; and because of symmetry, the two surface regions have identical widths and densities.

To find the equilibrium state in the SC-ISJM of a slab with $\nu$ atomic layers, we start from $\nu_2 = 1$, and solve simultaneously the equations

$$\left.\frac{\partial E(\nu, \nu_2; \bar{n}_1, \bar{n}_2)}{\partial \bar{n}_1}\right|_{\bar{n}_1 = \bar{n}_1^\dagger, \bar{n}_2 = \bar{n}_2^\dagger} = 0, \quad \left.\frac{\partial E(\nu, \nu_2; \bar{n}_1, \bar{n}_2)}{\partial \bar{n}_2}\right|_{\bar{n}_1 = \bar{n}_1^\dagger, \bar{n}_2 = \bar{n}_2^\dagger} = 0, \qquad (31)$$

to determine those $\bar{n}_1^\dagger$ and $\bar{n}_2^\dagger$ which minimize the total energy per electron. Then we change the value of $\nu_2$ and repeat the foregoing procedure until we find the value $\nu_2^\dagger$ for which the total energy becomes a global minimum. In this way, the surface and inner regions are determined self-consistently. For a given set of values $\nu$, $\nu_2$, $\bar{n}_1$, and $\bar{n}_2$, we expand the KS orbitals in terms of the eigenfunctions of a rectangular infinite potential well, as discussed in detail in reference [1]. In the variation process of $\bar{n}_1$ and $\bar{n}_2$, we use the fact that the surface area of the slab as well as the number of ions in the jellium slices remain constant [1]. This constraint leads to the relation between the width and the density of each region slice as

$$L_\alpha^B \bar{n}^B = L_\alpha \bar{n}_\alpha, \qquad (32)$$

where $L_\alpha^B \equiv \nu_\alpha d_{100}^B$ and $L_\alpha \equiv \nu_\alpha d_{100}$ are the width of the region $\alpha$ with the interlayer spacings $d_{100}^B$ and $d_{100}$, respectively; and the superscript B stands for Bulk. Using equation (32) and the definition of the relaxation of region $\alpha$ as $\Delta_\alpha \equiv (L_\alpha - L_\alpha^B)/L_\alpha^B$, the variation of total energy with respect to $\bar{n}_\alpha$ is related to its variation with respect to $\Delta_\alpha$ by

$$\frac{\partial E(\nu,\nu_2;\bar{n}_1,\bar{n}_2)}{\partial \bar{n}_\alpha} = -\frac{\bar{n}^B}{(\bar{n}_\alpha)^2} \frac{\partial E(\nu,\nu_2;\Delta_1,\Delta_2)}{\partial \Delta_\alpha}. \qquad (33)$$

The above equations enable us to determine, directly, the equilibrium relaxations of each region, $\Delta_\alpha^\dagger$, and the equilibrium energy, $E(\nu,\nu_2;\Delta_1^\dagger,\Delta_2^\dagger)$ for given values of $\nu$ and $\nu_2$.

**Results and discussion**

We have used the SC-ISJM to determine the equilibrium-state properties of $\nu$-layered slabs of Al, Na, and Cs for $3 \leq \nu \leq 10$. In these calculations, the bulk density parameters, $r_s^B \equiv (3/4\pi \bar{n}^B)^{1/3}$, are taken to be 2.07, 3.99, and 5.63 for Al, Na, and Cs, respectively. Consistent with these values, the bulk interlayer spacings for the (100) facets, $d_{100}^B$, of Al, Na, and Cs, in atomic units, are 3.82, 4.05, and 5.72, respectively. These interlayer spacings in units of their corresponding Fermi wavelengths of valence electrons are 0.56, 0.31, and 0.31, respectively.

| Element | $\nu$ | | | | | | | |
|---|---|---|---|---|---|---|---|---|
| | 3 | 4 | 5 | 6 | 7 | 8 | 9 | 10 |
| Al | 1 | 1 | 1 | 1 | 1 | 1 | 1 | 1 |
| Na | 1 | 1 | 1 | 1 | 1 | 1 | 1 | 1 |
| Cs | 1 | 1 | 2 | 1 | 2 | 1 | 2 | 2 |

**Table 1.** Values of $\nu_2^\dagger$ which correspond to the equilibrium states of $\nu$-layered slabs.

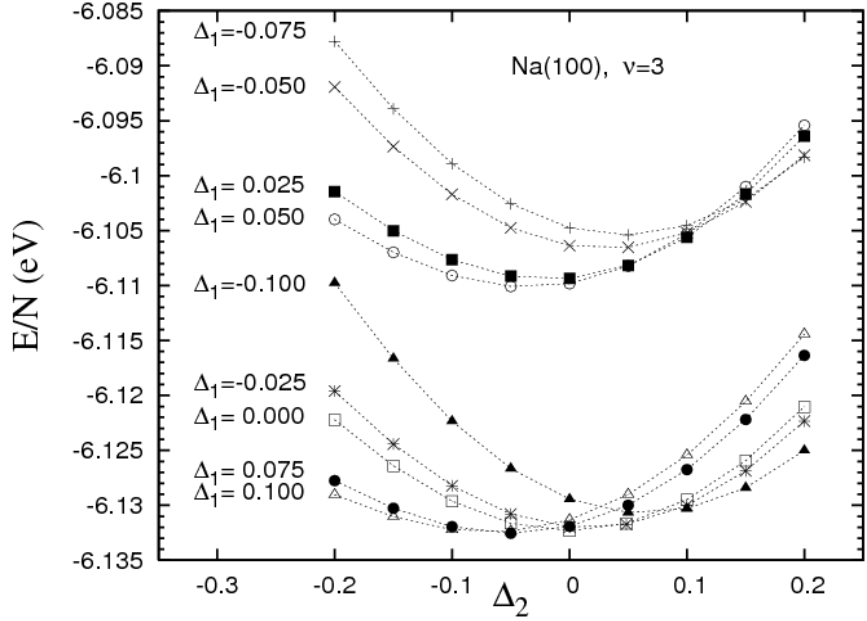

**Figure 3.** Variations of *E/N* with respect to $\Delta_2$ for different $\Delta_1$ values in a 3-layered Na slab.

In figure 3, we have plotted the variations of *E/N* of a 3-layered Na slab with respect to $\Delta_2$ for different values of $\Delta_1$. As is seen, the energy is much more sensitive to the values of $\Delta_1$ than the $\Delta_2$, so that we had to take a finer mesh for $\Delta_1$ around the equilibrium state. For the rough mesh points in figure 3, the values of $\Delta_1^\dagger$ and $\Delta_2^\dagger$ are seen to be as +0.075 and -0.05, respectively whereas, upon taking a finer mesh, we obtain more exact values of +0.058 and -0.038, as shown in figure 5(c)-(d). In this case of 3-layered slab, the $\nu_2 = 1$ is the only possible configuration and therefore, we have $\nu_2^\dagger = 1$. Similarly, for a 4-layered slab, the $\nu_2^\dagger = 1$ is the only possible configuration for the global-minimum state. However, for $\nu \geq 5$, the states with $\nu_2 = 2$ should also be considered. That is, for a $\nu$-layered slab, we have to consider all configurations for which the condition $1 \leq \nu_1 \leq \nu - 2$ is satisfied (see the appendix for different configurations).

Our calculations show that in the equilibrium states of the Al and Na slabs the surface regions contain only one atomic layer. That is, the global-minimum of the energies happen for $\nu_2^\dagger = 1$, whereas in some Cs slabs it occurs for $\nu_2^\dagger = 1$ and in some other Cs slabs for $\nu_2^\dagger = 2$. The calculated values of $\nu_2^\dagger$ are summarized in table 1.

In figure 4(a), we have plotted the energies $E(\nu, \nu_2; \Delta_1^\dagger, \Delta_2^\dagger)$ of the Al slabs as functions of $\nu$ for different values of $\nu_2$ (refer to appendix to see the evolution of the ground-state electron and jellium densities for different $\nu$ and $\nu_2$ values.) As is seen, the plot with $\nu_2 = 1$ lies below the other plots and therefore, we conclude that the surface region contains only one atomic layer, i.e., $v_2^\dagger = 1$. On the other hand, with increasing $v_2$, the energies approach the SC-SJM results. The $v_2 = 4$ plot which applies for $v \geq 9$ is excluded from the figures for better visualization of other parts. Similar plots for Na and Cs slabs are shown in figures 4(b) and 4(c), respectively. In the Na and Cs case, the energies for different $v_2$ values of a given $v$ are so close that they have overlapped in the corresponding figures. The differences are of orders $10^{-4}$ and $10^{-5}$ eV for Na and Cs, respectively. The overlapping data points in Na and Cs cases imply that the interaction energy between the electrons and the ions are not sensitive to the exact positions of the ions, in contrary to the Al case. The sensitivity of the Al energies to the exact ionic positions manifests itself in the relaxation behavior which will be discussed later in the following. The density plots in the appendix show the possible different ionic and electronic charge distributions (different $v_2$-values) for $v$-layered slabs $(3 \leq v \leq 10)$. As our results show, for Na and Cs, the energies of different possible configurations corresponding to different $v_2$-values of a given $v$-layered slab are the same (within $10^{-4}$ eV). In the case of Cs, we explain this behavior by the fact that the density rearrangements are quite small. However, the density rearrangements in the Na slabs are not so small while the energies remain unchanged. This behavior can be explained if the interaction energy of the electrons with the Na pseudo ions be vanishingly small. In fact it is the case because; the JM gives the value $r_s = 4.18$ for the bulk stability, which is approximately equal to that of the Na (See figure 1 of reference [11].)

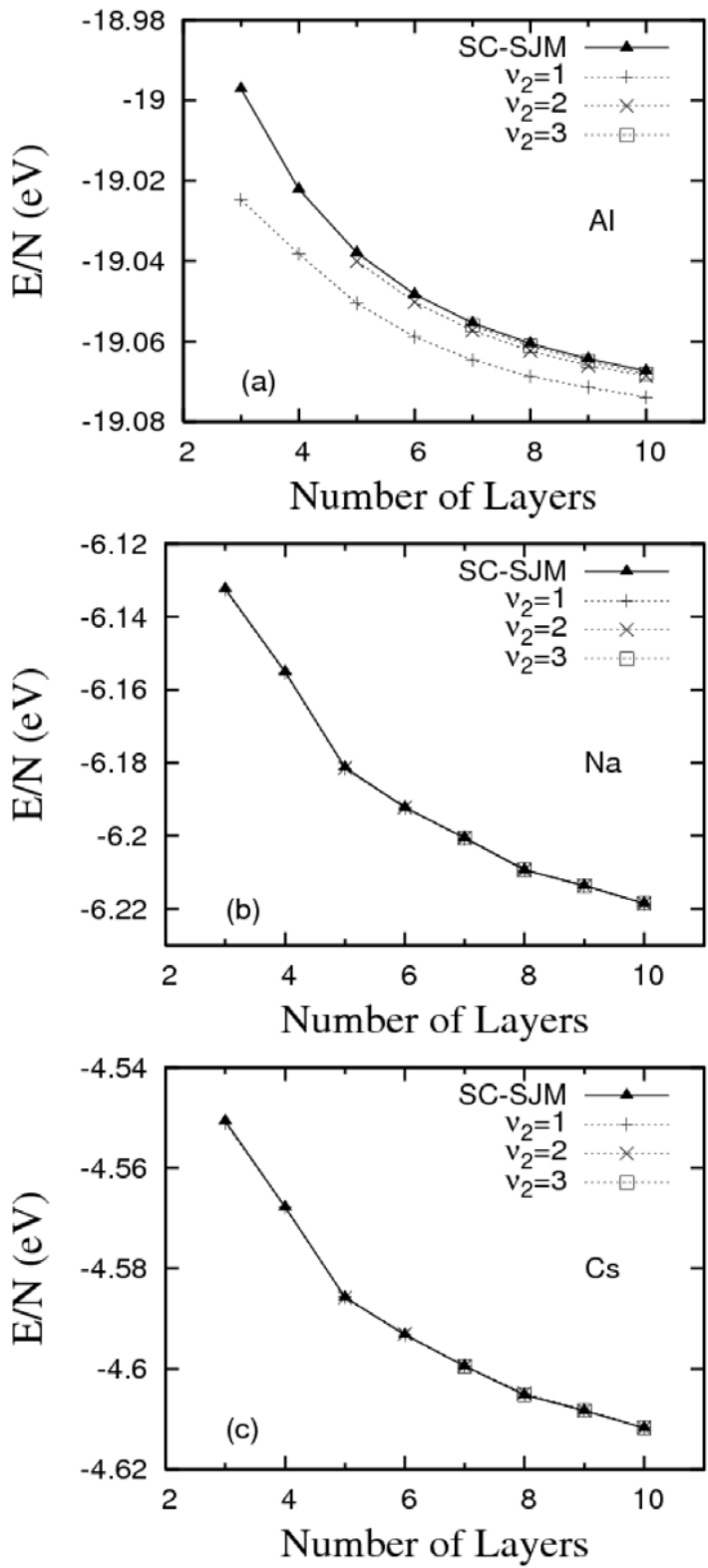

**Figure 4.** (a)-Total energies of Al slabs in the SC-ISJM as functions of the number of layers, $\nu$, for different surface parameters, $\nu_2$. For comparison, we have also included the SC-SJM results. Similarly, (b) and (c) correspond to Na and Cs slabs. As is seen, the overlapping data points in Na and Cs cases imply that the interaction energy between the electrons and the ions are not sensitive to the exact positions of the ions.

Inspecting the figures in the appendix, we observe that the rearrangements of both the jellium and the electron densities, compared to the SC-SJM results, are higher for Al slabs than for the Na and Cs ones. It means that even though the SC-SJM rearranges the densities with respect to those in the simple JM, to reduce the total force on the system (and thereby the total energy of the system), the local forces do not vanish. In the SC-ISJM, however, the rearrangements of the densities with respect to those in the SC-SJM, brings the system to a more stable state by reducing the local forces on the system. The density plots also imply that in the SC-SJM, because of lack of local charge-neutrality, the local forces on the surface layers of all slabs are higher than those on the interior layers. In the SC-ISJM, if we had considered more than two inequivalent regions, it would be possible to reduce the local forces even more, which on the other hand would increase the computational costs to the level of atomic simulations.

In figure 5(a)-(f), we have plotted the relaxations of surface and bulk regions separately as well as the overall relaxation of the Al, Na, and Cs slabs in their equilibrium states as functions of the number of atomic layers, $v$. In figure 5(a), it is observed that for Al the relaxation of the inner bulk region is as expansion whereas the relaxations of the surface regions are as contraction. The figure shows that the regional relaxations are larger for thinner slabs. However, the opposite relaxations of the inner and the surface regions lead to a relatively small overall relaxation. This is the new feature in the SC-ISJM which explains the relaxations at each region separately, whereas in the SC-SJM [1], it was possible to determine only the overall relaxation. The overall relaxation in the SC-ISJM for the equilibrium state of a $\nu$-layered slab is given by

$$\Delta^\dagger(\nu) = \nu^{-1}[(\nu - \nu_2^\dagger)\Delta_1^\dagger + 2\nu_2^\dagger \Delta_2^\dagger]. \qquad (34)$$

In figure 5(b), the overall relaxations for Al in the SC-ISJM and SC-SJM are compared and the comparison shows a similar trend with small differences in values. Both predict an overall contraction. This is in contrary to the first-principles all-electron or pseudopotential calculation results for the Al (100) slabs [12] (see also figure 4 of reference [1]) which show that the surface layers expand and the overall relaxations are as expansions.

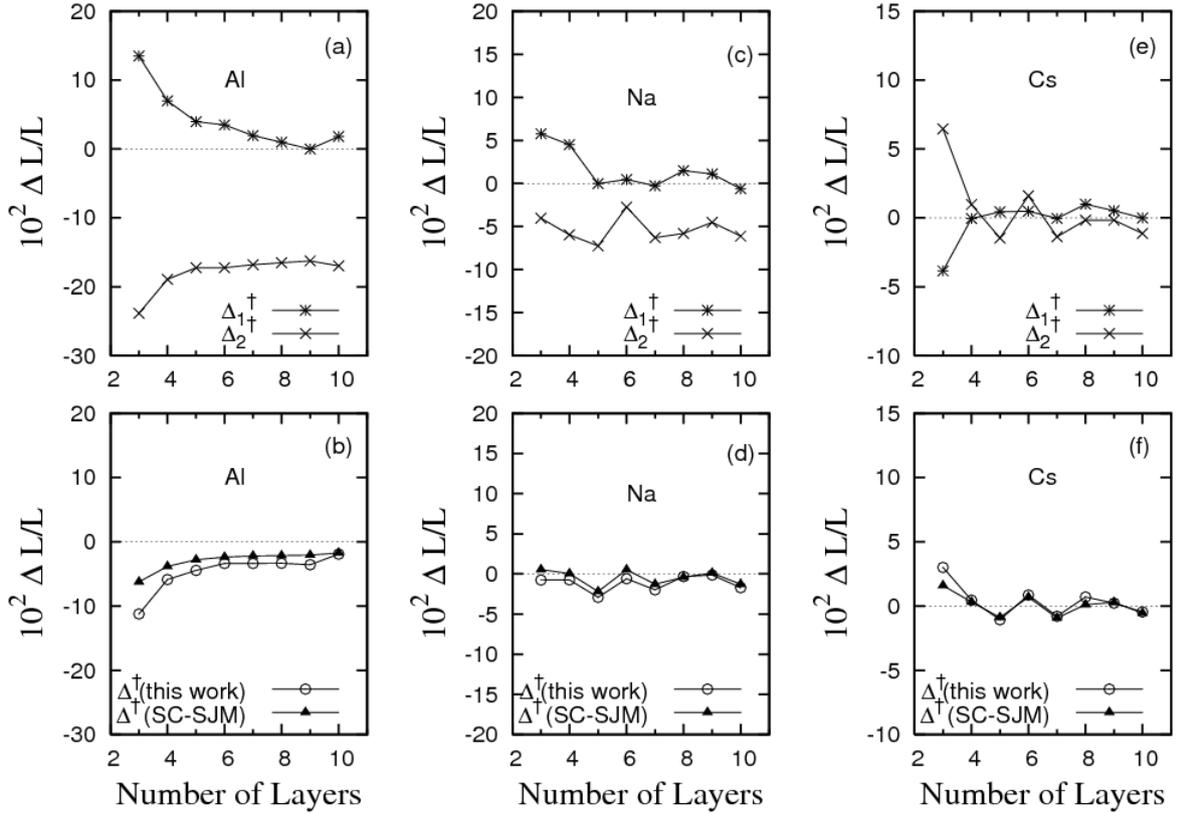

**Figure 5.** (a) Surface and bulk relaxations of Al slabs in their equilibrium states as functions of the number of atomic layers, $\nu$. (b) The overall relaxations of Al slabs in the SC-ISJM and SC-SJM as functions of $\nu$. Similarly, (c)-(d) correspond to Na slabs, and (e)-(f) correspond to Cs slabs.

The overall expansion of Al (100) slabs in the first-principles calculation results was explained [1] by resorting to the relation $2\lambda_F / d_{100} \gg 1$ and that the actual electron density must be much smaller than that in the free-electron model. To explain why the surface layers of the Al (100) slabs in the SC-ISJM undergo contraction (in contrast to the first-principles results which show expansions), our argument is the same. To this end, we have to concentrate on the interlayer spacings as well as the positions and widths of the Friedel peaks [13] in the electron densities presented in the appendix. In those cases where the Friedel peak is well contained inside the surface layer (see the results for all configurations $\nu_2 = 1, 2, 3, 4$), the maximum possible local charge neutrality happens when the surface layer contracts (or the jellium density increases). On the other hand, the surface layer expansion happens in those cases where the Friedel peak lies not well inside the surface region but near to the inner region, as in $Cs_3$, $Cs_4$, and $Cs_6$ [figure 5(e)]. Mathematically, the expansion of the surface layers can happen in those cases where the condition $\lambda_F / d_{100} >> 1$ is satisfied by the system [1]. However, for Al slabs this criterion is not satisfied when we consider the free-electron $r_s$ value with $z=3$. This means that to satisfy the criterion, we must have a larger Fermi wavelength which in turn implies that the effective valence density should be less than that of the free-electron model. To obtain more insights on why our model does not predict accurate behaviors for Al slabs, we have used the Ashcroft empty-core pseudopotential (with $r_c$ fixed demanding that the bulk stabilized jellium Al system becomes mechanically stable) to simulate Al (100) slabs in the first-principles pseudopotential calculations using the Quantum ESPRESSO code [14]. The results had shown that, even using the Ashcroft pseudopotential, the Al (100) slabs undergo expansions, in agreement with the results obtained using more sophisticated pseudopotentials. This result leads us to the conclusion that, in the Al case, the structural effects are not negligible [as we had seen in figure 4(a)], and therefore, some corrections to the "structureless" SC-ISJM have to be added.

The equilibrium relaxations of the Na slabs are shown in figure 5(c)-(d). As is observed in figure 5(c), the surface layers are contracted while the inner regions undergo expansions. On the other hand, in figure 5(d), we observe that the SC-SJM and the SC-ISJM results for the overall relaxations have good quantitative agreements; and except for the $Na_3$, $Na_4$, and $Na_6$ cases, the relaxations are of the same type. That is, for the exception cases the SC-SJM predicts expansions while the SC-ISJM predicts contractions. Compared to the Al slabs, the overall relaxations for the Na slabs are smaller.

In figure 5(e), the equilibrium-state relaxations of the surface and inner regions of the Cs slabs are presented. As is observed, the surface layers are expanded for $Cs_3$, $Cs_4$, and $Cs_6$, while the inner regions are expanded for $\nu \geq 5$. The overall relaxations which are presented in figure 5(f) show that they are smaller than those in the Na case.

The work functions of the Al slabs are presented in figure 6(a) as functions of the slab size for different surface parameters, $\nu_2$. The $\nu_2 = 1$ plot corresponds to the equilibrium state, as we discussed earlier. As is seen, the equilibrium-state work function has been decreased within about 5% compared to the SC-SJM results. The origin of the decrease in the work function in the SC-ISJM compared to the SC-SJM is that in the SC-ISJM, we have an inhomogeneous jellium background which allows larger overlaps between the electronic charge density at the surface region. The overlap is even larger when $\nu_2 = 1$. This in turn, reduces the surface dipole moments and thereby reduces the work function. This point had been also discussed in reference [1].

The work functions of Na and Cs slabs are presented in figures 6(b) and 6(c), respectively. As is seen, for Na and Cs, the SC-SJM and the SC-ISJM results are more or less the same. Referring to the figure 5(c) and the corresponding figures in the appendix, we explain this behavior to be due to insignificant changes in the local charge overlaps near the surface regions. As is observed from the density plots in the appendix, the behavior of the electronic densities in the Na slabs (in contrast to those in the Al slabs), are not much sensitive to the jellium deformations. Therefore, it is possible that the jellium densities near the surface regions vary in such a way that the surface dipole moments remain almost unchanged.

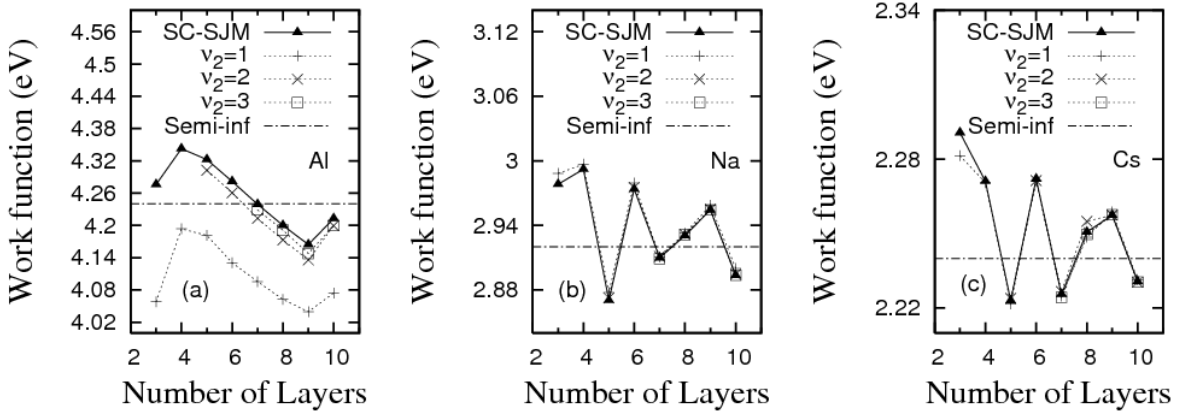

**Figure 6.** (a)-Work functions of Al slabs in the SC-ISJM as functions of the number of layers, $\nu$, for different surface parameters, $\nu_2$. The $\nu_2 = 1$ plot corresponds to the equilibrium state. For comparison, the SC-SJM and semi-infinite jellium [15, 16] results are included. Similarly, (b) and (c) correspond to Na and Cs slabs.

For the Cs slabs, from figure 5(e), we observe that the redistributions of the charges are not so significant and therefore the surface dipole moments remain almost unchanged. For comparison, the semi-infinite jellium results [15, 16] are also included in figures 6(a)-(c).

In the SC-ISJM, the surface energy of the slab is defined by

$$\sigma(\nu, \nu_2^\dagger; \Delta_1^\dagger, \Delta_2^\dagger) = \frac{1}{2A}[E(\nu, \nu_2^\dagger; \Delta_1^\dagger, \Delta_2^\dagger) - E(\nu, 0; 0, 0)], \qquad (35)$$

where, $A$ is the surface area, the first term in the brackets is the equilibrium energy of a $\nu$-layered slab in the SC-ISJM, and the second term is the SJM energy of the slab with bulk interlayer spacings, which is proportional to the number of layers, and can be calculated via linear fitting to the total energies [1]. In figure 7(a), we have presented the equilibrium SC-ISJM surface energies of the Al slabs as functions of the number of layers. The self-consistent surface energies for other surface parameters ($\nu_2 = 2, 3$) as well as the SC-SJM [1] and the semi-infinite jellium results [15, 16] are also included. The results for the equilibrium states have the lowest surface energies because the equilibrium SC-ISJM energies lie lower than the SC-SJM results while the values for the slabs with bulk spacings remain unchanged in the two schemes. The surface energies of the Na and Cs slabs are presented in figures 7(b) and 7(c), respectively. Since the energy changes, for Na and Cs, in the SC-SJM and SC-ISJM are not significant, we do not observe any significant changes in the surface energies.

**Conclusions**

We have formulated the SC-ISJM to explain the detailed variations of the interlayer spacings near the surfaces of the slabs of simple metals. However, since the next improvement over the SC-SJM can be obtained when we separate the surface region from the inner bulk region, we have applied the simplified version of the SC-ISJM to predict the relaxations as well as the other properties of the Al, Na, and Cs slabs. The number of atomic layers contained in the surface region is determined self-consistently from the global-minimization of the SC-ISJM energy of the slab. The results show that for the equilibrium states of Al and Na, the surface regions contain only one atomic layer while for some Cs slabs, it contains two atomic layers. The SC-ISJM provides the system the possibility to decrease local forces and thereby brings the system to a more stable state. The density plots show that, when the Friedel peak is well contained inside the surface region, that region is contracted. On the other hand, if that peak lies near to the inner region, there would be the possibility that the surface region expands. In other words, if the condition $\lambda_F / d_{100} >> 1$ is met, then it becomes possible to realize expansions, as in the Cs case. From this condition, and the fact that the first-principles results predict expansions for the surface layers of the Al (100) slabs whereas our SC-ISJM results predict contractions, we conclude that the actual electron densities near the surfaces of Al (100) slabs might be much less than those predicted by the SC-SJM. However, our atomic simulation with Ashcroft pseudopotential for Al (100) slabs confirmed that in Al, the electron-ion interaction energy is sensitive to the exact ionic positions and therefore, the "structureless" SC-SJM and SC-ISJM do not give accurate results for the properties that highly depend on the crystal structure. The work functions of Al slabs are significantly changed with respect to the SC-SJM values because in the SC-ISJM the surface dipole moments, which give rise to the work functions, are decreased. However, the decrease in the Na and Cs work functions are not significant because the values of the surface dipole moments have not been changed significantly. In the SC-ISJM, the surface energies decrease because the isolated slab energies decrease while the energies of the slabs with bulk interlayer spacings remain unchanged in the SC-SJM and SC-ISJM schemes.

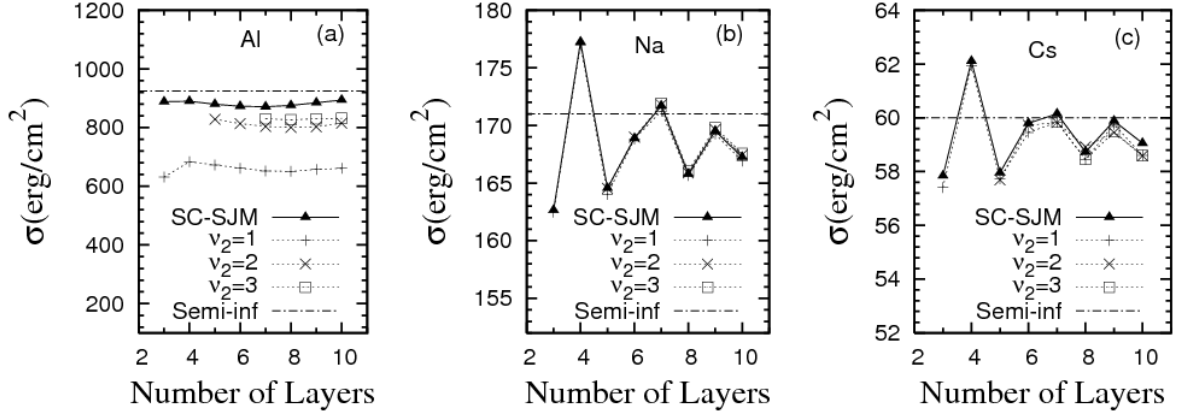

**Figure 7.** (a)-Surface energies of Al slabs in the SC-ISJM as functions of the number of layers, $v$, for different surface parameters, $v_2$. The $v_2 = 1$ plot corresponds to the equilibrium state. For comparison, the SC-SJM and semi-infinite jellium [15, 16] results are included. Similarly, (b) and (c) correspond to Na and Cs slabs.

**Acknowledgements**

M. P. would like to thank the useful comments of Paolo Giannozzi on using Ashcroft pseudopotential in the Quantum ESPRESSO code for simulating Al (100) slabs. This work is part of research program in NSTRI, Atomic Energy Organization of Iran.

**Appendix: Self-consistent electron and jellium densities**

In this section, the self-consistent electron and jellium densities are plotted as functions of the z-coordinates of the Cs, Na, and Al slabs. The subfigures at each column are arranged so that the bottom subfigure corresponds to SC-ISJM result for the state with configuration $(\nu, \nu_2 = 1; \Delta_1^\dagger, \Delta_2^\dagger)$, the next to the bottom subfigure corresponds to the SC-ISJM result for the state with configuration $(\nu, \nu_2 = 2; \Delta_1^\dagger, \Delta_2^\dagger)$, and so on; and the topmost subfigure corresponds to the self-consistent SC-SJM result. The densities are in units of the corresponding bulk densities of the valence electrons $\bar{n}^B$. The thin solid lines and the thick solid lines correspond to the jellium densities and the electron densities, respectively. For a given region in slab, if the jellium density lies above the dotted line, according to equation (32), that region is contracted and if it lies below the dotted line, that region is expanded. In the subfigure A1(a), the SC-SJM results for the 3-layered Cs slab show that the slab is slightly expanded. On the other hand, the corresponding SC-ISJM results in the subfigure A1(b) show that the outer layers are expanded whereas the inner layer contracted. Inspection of subfigures A1(c)-(d) reveals that the 3-layered Na slab undergoes a slight expansion in the SC-SJM, whereas in the SC-ISJM the outer layers contract while the inner layer expands. On the other hand, as shown in figures A1(e)-(f), the 3-layered Al slab is contracted in the SC-SJM whereas, in the corresponding SC-ISJM, the contraction of the outer layers and the expansion of the inner layer are higher than those in the Na case. The figures for different numbers of layers $\nu$ ($3 \leq \nu \leq 10$), and different numbers of surface layers $\nu_2$ show us, explicitly, the evolutions of the densities with respect to the sizes.

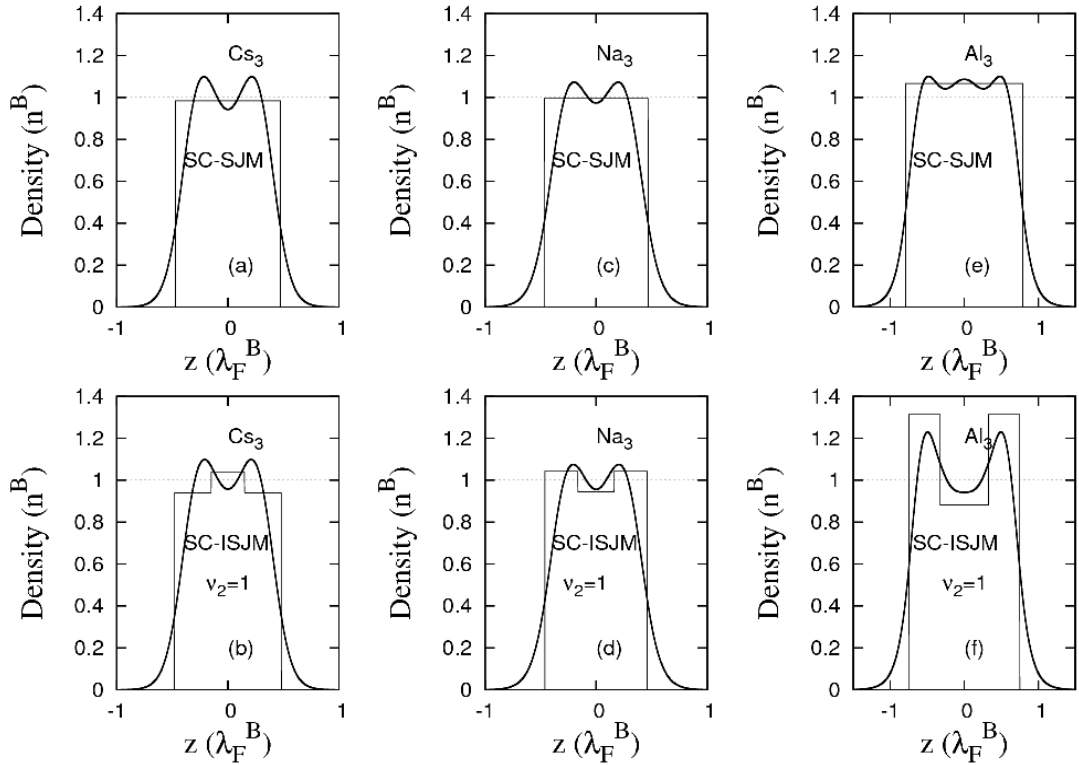

**Figure A1.** Self-consistent electron and jellium densities, in units of the bulk valence densities, of $\nu$-layered, with $\nu = 3$ Cs, Na, and Al slabs as functions of the $z$ coordinates, in units of the corresponding bulk Fermi wavelength of the valence electrons, $\lambda_F^B$. The thin solid and thick solid lines correspond to the jellium and electron densities, respectively; and the dotted line specifies the bulk density. The topmost subfigures correspond to the SC-SJM results, and the bottom subfigures correspond to the SC-ISJM results for surface regions containing one atomic layer, i. e., $\nu_2 = 1$.

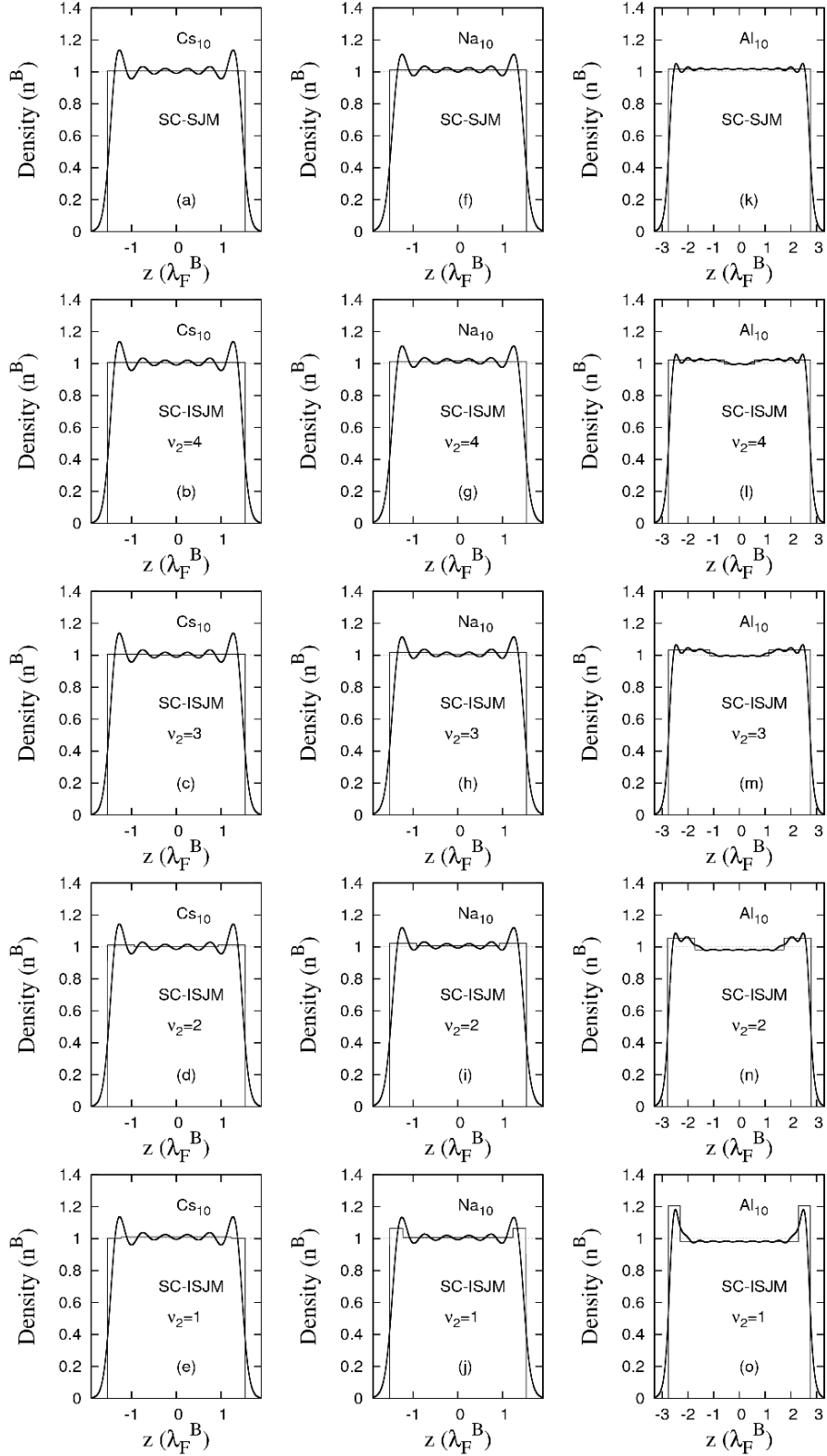

**Figure A2.** (a), (b), (c), (d),(e) correspond to the 10-layered Al slabs for SC-SJM, SC-ISJM with $v_2 = 4$, SC-ISJM with $v_2 = 3$, SC-ISJM with $v_2 = 2$, and SC-ISJM with $v_2 = 1$, respectively. It is clearly seen that in the SC-ISJM, the electron densities and jellium densities are so arranged that the local charge neutrality becomes greater than that in the SC-SJM.